\documentclass[12pt]{article}

\usepackage{amssymb,amsmath,epsfig}

\newcommand{\pd}{\partial}

\title{Energy Levels of ``Hydrogen Atom'' in Discrete Time Dynamics}
\date{}
\author{Andrei Khrennikov\thanks{Andrei.Khrennikov@msi.vxu.se} and Yaroslav Volovich\thanks{Yaroslav.Volovich@msi.vxu.se}\\
~\\
        International Center for Mathematical Modeling\\
        in Physics, Engineering and Cognitive science\\
        MSI, V\"axj\"o University, S-35195, Sweden
}

\begin{document}

\maketitle
\begin{abstract}
We analyze dynamical consequences of a conjecture that there exists
a fundamental (indivisible) quant of time. In particular we study
the problem of discrete energy levels of hydrogen atom. We are able
to reconstruct potential which in discrete time formalism leads to
energy levels of unperturbed hydrogen atom. We also consider linear
energy levels of quantum harmonic oscillator and show how they are
produced in the discrete time formalism. More generally, we show
that in discrete time formalism finite motion in central potential
leads to discrete energy spectrum, the property which is common for
quantum mechanical theory. Thus deterministic (but discrete time!)
dynamics is compatible with discrete energy levels.
\end{abstract}

\maketitle

\section{Introduction}

Discovery of discrete energy levels for atoms demonstrated that the
classical Newtonian model could not be used to describe this
phenomenon. Now days this phenomenon is described in the framework
of quantum mechanics. The advantages of the computational methods of
quantum mechanics are well known. They are widely used for
computation of energy levels not only of atoms, but also in
essentially more complicated situations. However, despite this
computational power, quantum mechanics induced many still
unsolvable phenomenological problems, see \cite{Heis}-\cite{Vax4}
for extended discussions. One of  distinguishing features of quantum
mechanics (at least in Copenhagen interpretation) is the
impossibility to provide realist deterministic  description of
quantum reality. In particular, here particles do not have well
defined trajectories.

The impossibility to use the deterministic evolutionary model looks
rather counterintuitive. Moreover, it  contradicts to many quantum
experiments where "trajectories" of particles are well observed,
e.g. in Wilson's camera. Of course, in modern quantum phenomenology
this problem is solved via the principle of complementarity.
Nevertheless, this strong deviation from our intuitive picture of
physical reality induces a rather general opinion that quantum
mechanics has a lot of mysteries \cite{FeyGib}.

This unsatisfactory status of quantum mechanics induces new and new
attempts to create new models that would be closer to our physical
intuition. We just mention one model, Bohmian mechanics
\cite{BH,Hol}.

Last years there were performed intensive investigations to
reconsider probabilistic foundations of quantum mechanics, see e.g.
\cite{Vax3,Vax4}. These investigations are of the great importance.
It is well known that "quantum probability" differs strongly from
"classical probability". Typically, see e.g. \cite{FeyGib}, it is
pointed out that quantum randomness is irreducible and fundamental
(in the opposite to classical randomness that could be reduced to
e.g. randomness of initial conditions). This irreducible  quantum
randomness is deeply connected with the impossibility of realist
deterministic models that would reproduce quantum probabilities. It
seems to be impossible to imagine that quantum particles have
deterministic trajectories, since the existence  of  trajectories
should imply classical probabilistic rules. But these rules are
violated (for example, in the two-slit experiment \cite{FeyGib},
see also investigations on EPR-Bohm-Bell consideration and chameleon effect \cite{Bell,Acc,AIR}).

In papers \cite{Kh,Kh5,Kh6,Kh8} there was developed the contextual
approach to quantum probabilities that might be used to explain the
origin of quantum randomness in the classical (but contextual!)
probabilistic framework. Therefore the probabilistic constraint to
create realist deterministic models need not be taken into
account\footnote{Of course, there are also Bell`s inequality
arguments. But there arguments are strongly coupled to nonlocality.
And this problem is far from problems considered in the present
paper.}.

What kind of realist deterministic models could be created for
quantum phenomena?

As we have already mentioned, we could not directly use Newtonian
mechanics even for the simplest quantum system -- the hydrogen atom.
Thus Newton equations should be changed. The most straightforward
idea is to change classical forces (in the case of atom Coulomb's
law) and find new forces, quantum forces. Such forces should "drive"
particles along trajectories that reproduce quantum data. There is a
rather common view point that one of successful realizations of such
a program is given by Bohmian mechanics. However, this problem is
not simple, since Bohmian mechanics is, in fact, not mechanics, but
a field model. There exists an additional field equation for quantum
potential. We do not know any model that would give the realization
of the discussed program of "modernization" of Newtonian mechanics.

In the present paper we consider discrete time Newtonian model. This
is a kind of classical physical model (in particular, realist
deterministic). The only difference is discreteness of time. Thus we
use the classical force -- interaction picture, but the discrete
time version of Newton's second law. One of the main advantages of
this model is its simplicity. We do not change phenomenology of
classical physics (position, velocity, force-interaction,
trajectories). We only change the mathematical representation of
time.

On the other hand, discreteness is the main distinguishing feature
of quantum physics. In fact, M. Planck and A. Einstein obtained
Wien's law simply by assuming discreteness of energy. However, this
discreteness approach was not developed further to get a formalism
based only on the discreteness postulate. Discreteness of quantum
observables was reproduced by using an advanced mathematical
formalism based on the representation of the observables in the
complex Hilbert space. As we have already mentioned, the use of this
formalism (despite its computational advantages)  induced many
phenomenological problems. We are trying to modify classical physics
by starting with one natural postulate:

\medskip
\textbf{\underline{TD:}~~``Time is discrete''.}

\medskip
Here, in particular, continuous Newton equations (differential
equations) are just approximations of real equations of motions,
namely, Newton's difference equations\footnote{We reverse the modern
viewpoint on the description of physical reality. Not (discrete)
difference equations are used to approximate (continuous)
differential equations, but the inverse!}. In the present paper we
use the discrete time model for the hydrogen atom. The discrete-time
postulate implies discrete orbits and energy levels. At the same
time we have deterministic motion along orbits. In the limit we get
continuous motion along circular orbit (a kind of Bohr's
correspondence principle).

In the present paper we do not try to develop some statistical
theory (a kind of Born approach). Our investigation has some
similarities with the original paper of W.~Heisenberg (that was not
statistical one), \cite{Heis,Heis2}. We hope that starting with only
discrete time postulate we would be able to develop simpler
formalism to calculate spectra of quantum observables - without
using noncommutative calculus and without a cardinal change of the
phenomenology of classical physics.

It is interesting to point that there is some similarity between our
approach and G. 't Hooft's approach \cite{Hooft1,Hooft2,Hooft3}
where a general scheme was proposed that maps states of quantum
field system to the states of a completely deterministic field
model. Although in this note we do not consider the field-particle
duality it would be interesting to study this problem.

Some statistical consequences of the postulate (TD) were
investigated in our previous papers \cite{KV1,KV2,KV3}.

\section{Discrete Time Dynamics}
\label{sec:ddyn}

In classical mechanics a dynamical function
$A=A(p,q)$ (here $p$ and $q$ are momenta and coordinates
of the system) evolves according to the following well known equation
\cite{Dir}
\begin{equation}
\label{dyneq} D_t A=\{A,H\}
\end{equation}
where $H=H(p,q)$ is a Hamiltonian of the system and in the right
hand side is a Poisson bracket, which could be presented as
\begin{equation}
\label{cposs} \{A,B\}=\frac{\pd A}{\pd q}\frac{\pd B}{\pd p} -
\frac{\pd A}{\pd p}\frac{\pd B}{\pd q}
\end{equation}
The left hand side of (\ref{dyneq}) contains a continuous time derivative
$$
D_t A = \frac{dA}{dt}
$$
As it was mentioned earlier we are interested in construction
dynamics with discrete time. This is done with the help of
\textit{discrete derivative} which is postulated to be
$$
D^{(\tau)}_t A = \frac{1}{\tau}[A(t+\tau)-A(t)],
$$
where $\tau$ is the discreteness parameter. This parameter is finite
and is treated in the same way as Plank constant in quantum
mechanical formalism. In particular if $\tau$ is small relative to
dimensions of the system then classical approximation with
continuous derivative might work well (although this could not be
the case all the time in the same sense as there are examples when
quantum formalism is reasonable even for macroscopic systems, for
example in superfluidity).

Summarizing, the discrete time dynamical equation is postulated to
be
\begin{equation}
\label{dyneq-post} D^{(\tau)}_t A=\{A,H\},
\end{equation}
where $A(p,q)$ is a real-valued function of real-valued dynamical
variables and in the right hand side there is classical Poisson
bracket (\ref{cposs}). The equation (\ref{dyneq-post}) could be
solved in the sense that we can write
\begin{equation}
\label{dyn-sol} A(t+\tau)=A(t)+\tau \{A,H\}
\end{equation}
thus providing the evolution of any dynamical function $A=A(p,q)$.

Note that in our model the coordinate space is continuous.

\section{Motion in Central Potential}

\begin{figure}
\centering \epsfig{file=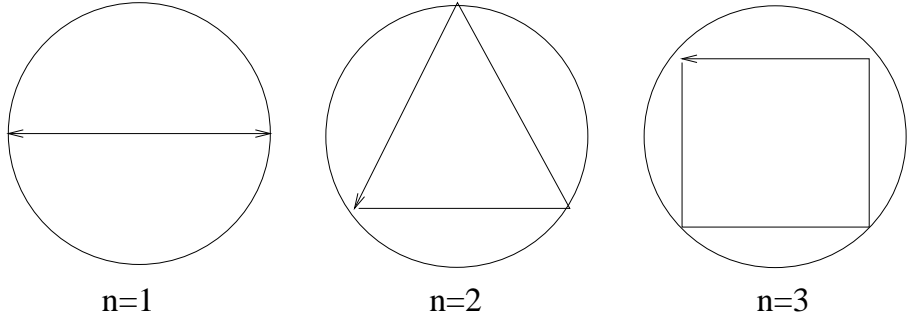,width=12cm} \caption{First three
trajectories.} \label{fig:traj}
\end{figure}

Here we will study the properties of motion in central potential
$U=U(r)$ in discrete time formalism. As we will see discreteness of
time enriches mechanics with some new properties which are usually
thought as having quantum nature. In particular as it will be shown
below in discrete time mechanics stationary orbits (i.e. finite
motion) have discrete energy spectrum. We point that the phase space
is assumed here to be a continuous real manifold.

Following the general approach described in the previous section we
start from the classical Hamiltonian and then write the dynamical
equations. In polar coordinates $(r,\varphi)$ the Hamiltonian of the
system with mass $m$ in central potential $U(r)$ is given by
\begin{equation}
\label{HamiltU}
H(r,p_r,\varphi,p_\varphi)=\frac{p_r^2}{2m}+\frac{p_\varphi^2}{2mr^2}+U(r),
\end{equation}
where $p_r$ and $p_\varphi$ denote momenta corresponding to $r$ and
$\varphi$ -- radial and angular coordinates respectively. Using
(\ref{dyn-sol}) let us write the dynamical equations. We obtain
\begin{align}
\label{r-dyn}
r(t+\tau)&=r(t)+\tau \frac{p_r}{m}\\
\label{pr-dyn}
p_r(t+\tau)&=p_r(t)+\tau \left(\frac{p_\varphi^2}{mr^3}-\frac{\pd U}{\pd r}\right)\\
\label{phi-dyn}
\varphi(t+\tau)&=\varphi(t)+\tau \frac{p_\varphi}{mr^2}\\
\label{pphi-dyn} p_\varphi(t+\tau)&=p_\varphi(t)
\end{align}
The equation (\ref{pphi-dyn}) corresponds to angular momentum
conservation in central field -- this is a direct analog of the
angular momentum conservation law in classical (continuous time)
dynamics and is a consequence of the fact that our Hamiltonian does
not depend on $\varphi$, (it is a so called \textit{cyclic
variable}).

Let us limit ourself to circular stationary periodic orbits. In this
case we should have $r(t+\tau)=r(t)$ and thus using (\ref{r-dyn}) we
see that the radial momentum should be zero,
\begin{equation}
\label{pr-zero} p_r(t)=0
\end{equation}
From (\ref{pr-dyn}) and (\ref{pr-zero}) we obtain the following
condition
\begin{equation}
\label{pu} \frac{p_\varphi^2}{mr^3}=\frac{\pd U}{\pd r}
\end{equation}
Let us finally come to the angular coordinate $\varphi$. For stable
motion the following \textit{periodicity} condition should be
satisfied (Fig. \ref{fig:traj})
\begin{equation}
\label{percd} \varphi(n\tau)=\varphi(0)+2\pi,
\end{equation}
where $n=1,2,\ldots$ (note that discreteness of $n$ is a consequence
of discreteness of time). Using (\ref{phi-dyn}) and (\ref{percd}) we
get
$$
n\tau\frac{p_\varphi}{mr^2}=2\pi,
$$
or
\begin{equation}
\label{p-phi} p_\varphi=\frac{2\pi m r^2}{n\tau}
\end{equation}
From (\ref{pu}) and (\ref{p-phi}) we get the following equation for
the radius of the $n$-th orbit
\begin{equation}
\label{rn} \frac{4\pi^2 m r}{n^2\tau^2}=\frac{\pd U}{\pd r}%,~~n=1,2,\ldots
\end{equation}
If potential $U(r)$ is known then from equation (\ref{rn}) we can
find $r=r_n$. If we consider physical potentials, i.e. potentials for which
the force, $-U'(r)$, is smooth, negative, and is strictly monotonically decreasing in absolute value as
$r$ grows, vanishing on infinity, then solution of (\ref{rn}) always exists and unique.
Now upon substituting $r_n$ to the Hamiltonian (\ref{HamiltU}) we obtain energy levels $E_n$.

For circular periodic orbits the original Hamiltonian
(\ref{HamiltU}) due to (\ref{pr-zero}) and (\ref{pu}) simplifies to
the following form
$$
H=\frac{1}{2}r\frac{\pd U}{\pd r} + U(r)
$$
Thus the expression for energy $E_n$ of the $n$-th stable periodic
orbit in terms of its radius $r_n$ is given by
\begin{equation}
\label{En} E_n=\frac{1}{2}r_n\left.\frac{\pd U}{\pd
r}\right|_{r=r_n}+U(r_n),~~n=1,2,\ldots
\end{equation}
As we see if potential $U(r)$ allows stationary periodic motion
and equation (\ref{rn}) has unique positive solutions $r_n$ then the
energy spectrum is discrete. This situation is directly analogous to
quantum mechanics where for finite motion we might expect discrete
energy levels.

\section{Energy Levels of Hydrogen Atom}
\label{sec:hydr}

Our task in this section is to study whether the discrete time
dynamics in central field described can lead to energy levels of
hydrogen atom. We treat energy levels as given (measured) quantities
and our task is to find the corresponding potential. We restrict
ourself to the simplest case when the atom is unperturbed by
external electric or magnetic fields and thus currently we do not
study splitting of the energy levels (note that in order to observe
``degenerate'' levels in quantum mechanical treatment one needs to
somehow perturb the system in such a way that the levels split).

We start from the following energy spectrum for hydrogen atom
\cite{Bohm}
\begin{equation}
\label{Enatom} E_n=-\frac{\gamma}{n^2},~~n=1,2,\ldots,
\end{equation}
where $\gamma\approx 13.6~eV$ is ionization energy of the hydrogen
atom. The task is to find such $U=U(r)$ that leads to the spectrum
(\ref{Enatom}). Using (\ref{rn}), (\ref{En}), and (\ref{Enatom}) we
get
\begin{equation}
\label{Enrna} \frac{1}{2}\xi
\frac{r_n^2}{n^2}+U(r_n)=-\frac{\gamma}{n^2},
\end{equation}
where the constant $\xi$ is given by (note that $\xi$ depends on
discreteness parameter $\tau$)
\begin{equation}
\label{xidef} \xi=\frac{4\pi^2m}{\tau^2}
\end{equation}
and $n=1,2,\ldots$ Let us rewrite equation (\ref{Enrna}) in the
following form
\begin{equation}
\label{Enrn2} U(r_n)=-\frac{1}{n^2}\left(\frac{1}{2}\xi
r_n^2+\gamma\right)
\end{equation}
We want to find $U(r)$. The idea is to obtain dependence of
$r_n=f(n)$ on $n$ and then inverting it substitute $n=f^{-1}(r_n)$
in (\ref{Enrn2}), here $f^{-1}$ denotes function inverse to $f$. As
a result of this procedure we will get rid of explicit dependence of
the right hand side of (\ref{Enrn2}) on $n$, it will depend only on
$r_n$. Then we interpolate the result for any $r\geqslant 0$. The
resulting $U(r)$ we check by substitution to (\ref{rn})-(\ref{En}).

\begin{figure}
\centering \epsfig{file=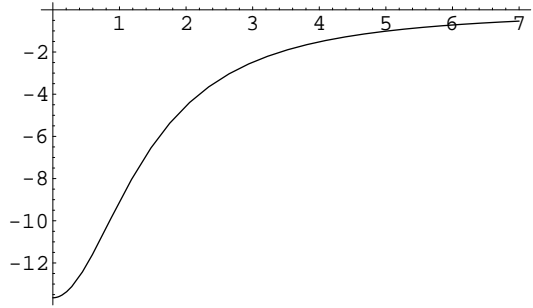,width=7cm} \caption{The shape of
potential (\ref{U-atom}) which leads to energy levels of the
hydrogen atom (we put $\beta=\xi=\gamma=1$).} \label{fig:U-atom}
\end{figure}

Let us proceeding as described above. First, we have to find
dependence of $r_n$ on $n$. To do this let us assume that $n$ is
continuous and take the derivative in $n$ of both sides of
(\ref{Enrn2}). This gives
\begin{equation}
\label{Ueq11} \left.\frac{\pd U}{\pd r}\right|_{r=r_n}\frac{d r_n}{d
n}= \frac{\xi r_n^2}{n^3}-\frac{\xi r_n
r_n^{\,\prime}}{n^2}+\frac{2\gamma}{n^3}
\end{equation}
Now, from (\ref{rn}) we find
\begin{equation}
\label{Ueq12} \left.\frac{\pd U}{\pd
r}\right|_{r=r_n}=\xi\frac{r_n}{n^2}
\end{equation}
Substituting (\ref{Ueq12}) to (\ref{Ueq11}) we obtain the following
differential equation for $r_n$
\begin{equation}
\label{rnatom} 2\xi n r_n r_n^{\,\prime} - \xi r_n^2 - 2\gamma=0
\end{equation}
Solving (\ref{rnatom}) and taking into consideration only positive
solution we obtain
\begin{equation}
\label{rnsolat}
r_n=\sqrt{\frac{-2\gamma+e^{2\beta\xi}n}{\xi}},
\end{equation}
where the constant $\beta$ is due to integration. Inverting
(\ref{rnsolat}) we get
\begin{equation}
\label{n-sol} n=e^{-2\beta\xi}(2\gamma+r_n^2\xi)
\end{equation}
Substituting (\ref{n-sol}) to (\ref{Enrn2}) we obtain
$$
U(r_n)=-\frac{e^{4\beta\xi}}{4\gamma+2r_n^2\xi }
$$
or performing interpolation, i.e. putting $r$ in place of $r_n$ we
finally get
\begin{equation}
\label{U-atom} U(r)=-\frac{e^{4\beta\xi}}{4\gamma+2r^2\xi }
\end{equation}
Now, substituting (\ref{U-atom}) to (\ref{rn})-(\ref{En}) we come to
the expected energy levels (\ref{Enatom}) of hydrogen atom.

The form of the potential (\ref{U-atom}) is presented on
Fig.\ref{fig:U-atom}. It is interesting to note that unlike
Coulomb's potential it is nonsingular at $r=0$.

\section{Spectrum of Harmonic Oscillator}
\label{sec:osc}

Procedure described in the previous section could be used to obtain
potentials corresponding to arbitrary energy spectrum. Here we
deduce the potential which results in linear energy levels of the
homogeneous two dimensional quantum harmonic oscillator.

In quantum mechanics homogeneous two dimensional harmonic oscillator
is a system described by the potential
\begin{equation}
\label{Vharm} U(r)=\frac{1}{2}m\omega^2 r^2,
\end{equation}
Solving the Schr\"odinger equation in potential (\ref{Vharm})
results in the following energy spectrum
\begin{equation}
\label{ELosc}
E_\Lambda=\hbar\omega(\Lambda+1),~~\Lambda=0,1,\ldots
\end{equation}
The simplest way to get this relation is to note that we deal with
two uncoupled oscillators (indeed, if $x$ and $y$ are Cartesian coordinates
then equation (\ref{Vharm}) takes the form $U=\frac{1}{2}m\omega^2(x^2+y^2)$),
each having energy
$E_k=\hbar\omega(k+\frac{1}{2})$, then the total energy is just the
sum of two such terms and we get (\ref{ELosc}).

Let us first rewrite (\ref{ELosc}) in terms of $n=1,2,\ldots$, we
have $n=\Lambda+1$ and thus
\begin{equation}
\label{Enosc}
E_n=\alpha n,~~n=1,2,\ldots,
\end{equation}
where $\alpha=\hbar\omega$. Our task is to find potential $U=U(r)$
which result in energy levels (\ref{Enosc}). Proceeding as in the
previous section we assume that $n$ is continuous and write the
differential equation
\begin{equation}
\label{Enalpha}
\frac{dE_n}{dn}=\alpha
\end{equation}
Now, using (\ref{rn}) and (\ref{En}) equation (\ref{Enalpha})
reduces to
\begin{equation}
\label{rnosc}
2\xi \frac{r_n r_n^{\,\prime}}{n^2}-\xi\frac{r_n^2}{n^3} = \alpha,
\end{equation}
where $\xi$ is given by (\ref{xidef}). The positive solution of
(\ref{rnosc}) is given by
\begin{equation}
\label{rnOSCsol}
r_n=\sqrt{\beta n + \frac{\alpha n^3}{2\xi}},
\end{equation}
where $\beta$ is a constant due to integration. This leads us to a
cubic equation in terms of $n$, which has one real solution
$$
n=\frac{\mathcal{V}(r_n)}{3\alpha}-\frac{2\beta\xi}{\mathcal{V}(r_n)},
$$
where $\mathcal{V}(r)$ is given by
$$
\mathcal{V}(r)=3^{1/3}\left(9r^2\alpha^2\xi+\sqrt{3}\sqrt{27r^4\alpha^4\xi^2+8\alpha^3\beta^3\xi^3}\right)^{1/3}
$$
We finally get the expression for potential
$$
U(r)=\frac{\mathcal{V}(r)}{3} -
  \frac{2\alpha\beta\xi}
   {\mathcal{V}(r)} -
  \frac{9r^2\,{\mathcal{V}(r)}^2\,
     {\alpha }^2\,\xi }{2\,
     {\left( {\mathcal{V}(r)}^2 -
         6\alpha\beta\xi\right)
         }^2}
$$
If we put $\beta=0$ in (\ref{rnOSCsol}) the potential takes simple
form
\begin{equation}
\label{U-osc} U(r)=\frac{3}{2}
\left(\frac{r^2\alpha^2\xi}{4}\right)^{1/3}
\end{equation}
Substituting (\ref{U-osc}) to (\ref{rn})-(\ref{En}) we get correct
energy levels (\ref{Enosc}). It is interesting to provide the
expression (\ref{U-osc}) when all constants are substituted, we have
$$
U(r)=\frac{3}{2}
\left(r\,\frac{\hbar\omega\pi\sqrt{m}}{\tau}\right)^{2/3}
$$

\section{General Case of Arbitrary Spectrum}
\label{sec:arb}

One can find explicit relations for the radii of the orbits for
arbitrary spectrum $E_n$. Indeed, From (\ref{rn}) and (\ref{En}) we
want to find $U$ as a function of $r$ in terms of a given energy
spectrum $E_n$. As we did above we assume the continuity of the parameter
$n$ and take derivative of both parts of (\ref{En}) in $n$, we have
$$
E^{\,\prime}_n=
r^{\,\prime}_n\frac{\xi r_n}{n^2}
-\frac{\xi r_n^2}{n^3}
+r^{\,\prime}_n\left.\frac{\pd U}{\pd r}\right|_{r=r_n},
$$
where prime denotes the derivative in $n$. The use of (\ref{rn})
allows us to get a differential equation for $r_n$ in terms of only
known quantities. We have
$$
2r_n r^{\,\prime}_n-\frac{1}{n}r_n^2=\frac{n^2}{\xi}E^{\,\prime}_n
$$
Introducing a new variable $\rho=r^2$ we obtain a linear
differential equation which could be rewritten as
$$
\frac{d}{dn}\left(\frac{\rho_n}{n}\right)=\frac{1}{\xi}n\,E^{\,\prime}_n
$$
which can be integrated to obtain
\begin{equation}
\label{nr}
r_n=\sqrt{\frac{1}{\xi}n \left(nE_n - E_1 - \int_{1}^n E_k dk + \epsilon \right)}
\end{equation}
Equation (\ref{nr}) expresses $n$-th radius in terms of $n$, i.e. it
has the form $r=f(n)$ now if we invert it we relate $n$ in terms of
$r$, $n=f^{-1}(r)$, which if substituted to (\ref{En}) gives an
equation for $U$ in terms of $r$ only (actually in terms of $r_n$,
but we perform interpolation effectively ignoring the fact that the
relation strictly holds only for orbit radii).

Note that in (\ref{nr}) we have a constant $\epsilon$ (having units of energy)
arising due to the integration, this means that we have a set of potentials
resulting in the same energy spectrum. As we see from (\ref{nr}) the constant
$\epsilon$ could be determined if for example the smallest radius $r_1$ is known.
In sections \ref{sec:hydr} and \ref{sec:osc} the situation was the same resulting in the
constant $\beta$ (see (\ref{rnsolat}) and (\ref{rnOSCsol})).
Since $\epsilon$ is more interesting from the point of view of physical interpretation we
provide the expressions for $\beta$ in terms of $\epsilon$. For the case of
energy levels of hydrogen atom we have (see (\ref{rnsolat}))
$$
\beta_{hydr}=\frac{1}{2\xi}\ln(\epsilon+\gamma)
$$
and for the case of harmonic oscillator we have (see (\ref{rnOSCsol}))
$$
\beta_{osc}=\frac{\epsilon - \frac{1}{2}\alpha}{\xi}
$$

\section{Energy Spectrum in Various Potentials}

As we already seen, for a given potential it is straightforward to
compute corresponding energy levels. Indeed, from (\ref{rn}) we find
$r=r_n$ and upon substitution to (\ref{En}) we get $E_n$. Like in
quantum mechanics, potential $U=U(r)$ should be attractive and
strong enough to result in finite motion. Below we consider several
common central potentials which result in rather simple expressions
for energy spectrum. In what follows the constant $\xi$ is given by
(\ref{xidef}), note that it depends on time discreteness parameter
as $1/\tau^2$.

\noindent a). Coulomb potential
$$
U(r)=-\frac{\alpha}{r},~~ r_n=n^{2/3}\left( \frac{\alpha}{\xi}
\right)^{1/3},~~
E_n=-\frac{1}{2n^{2/3}}\left(\alpha^2\xi\right)^{1/3}
$$
Note that energy spectrum is different from the $-\gamma/n^2$
spectrum obtained in quantum mechanics for this potential (see
section on energy levels of hydrogen atom for detailed discussion).

\vspace{0.1cm} \noindent b). Linear potential
$$
U(r)=\alpha r,~~ r_n=\frac{n^2\alpha}{\xi},~~
E_n=\frac{3n^2\alpha^2}{2\xi}
$$

\vspace{0.1cm} \noindent c). Logarithmic potential
$$
U(r)=\alpha\ln r,~~ r_n=n\sqrt{\frac{\alpha}{\xi}},~~
E_n=\alpha\left[\frac{1}{2}+\ln\left(n\sqrt{\frac{\alpha}{\xi}}\right)\right]
$$
For potentials (a), (b), and (c) $r_n>0$ if $\alpha>0$ in all three
cases it corresponds to the attraction field.

\vspace{0.1cm} \noindent d). Polynomial potential
\begin{equation}
\label{polyU}
U(r)=\alpha r^\sigma,~~ r_n=\left(\frac{n^2
\alpha\sigma}{\xi}\right)^{\frac{1}{2-\sigma}},~~
E_n=\frac{1}{2}\alpha (2+\sigma)
\left(\frac{n^2\alpha\sigma}{\xi}\right)^{\frac{\sigma}{2-\sigma}}
\end{equation}
This case generalizes cases (a) and (b) described above, although
because of importance of these potentials we provided corresponding
expressions explicitly. Note that if we make $\sigma$ in (\ref{polyU}) satisfy the following equation
$$
\frac{2\sigma}{2-\sigma}=1,
$$
i.e. $\sigma=2/3$ then we get the linear energy spectrum $E_n\sim n$
for the 2D quantum harmonic oscillator (see previous section for
detailed discussion).

\section{Discussion and Conclusion}

In this paper we have shown that discrete time formalism leads to
some distinguishable properties of micro-observables that are used
to be described with quantum mechanics. In particular it was shown
that finite motion results in discrete energy spectrum. Of the main
interest in this paper is discrete energy levels of hydrogen atom.
We have shown that for unperturbed hydrogen atom the discrete time
formalism is able to give correct energy spectrum, more precisely we
have reconstructed the corresponding ``micro''-potential. Here we
did not consider Stark or Zeeman effects for hydrogen atom, it would
be interesting to study them from the point of view of discrete time
formalism.

As we have seen in above the discrete time model requires potentials
which are different from QM potentials. One may argue this as a
disadvantage of the model. We pay attention that there are no
reasons to expect to reproduce QM by using the standard classical
potentials; Bohr, Zommerfeld, Heisenberg and many others tried to do
this, but they did not succeed. D. Bohm developed  a new model
\cite{BH} in which quantum mechanics can be reproduced on the
classical basis, but, of course, the classical potential could not
be preserved -- it is perturbed by the quantum potential. And the
latter looks not so natural from the classcial viewpoint, see e.g.
the quantum potential for the two slit experiment in \cite{BH}. The
discrete time model has an analogy with Bohmian mechanics -- it
tries to reproduce QM by changing potentials. But there is of course
the fundamental difference: the only postulate that is used in the
proposed approach is that there  exists a quant of time $\tau$.

Another interesting point is that one might expect that our
dynamical equations are essentially difference equations which might
produce discrete spectrum. This is not correct -- we recall that in
our model only time is discrete, but space is still continuous.

There is still an open problem of the quantative value of the
discreteness parameter $\tau$. One might speculate its relation with
Plank time constant \cite{Plank,QuantGrav} -- the smallest
measurable time interval in ordinary QM and gravity -- which is
quantatively given by
$$
t_{Pl}=\sqrt{\frac{\hbar G}{c^5}}\approx 5.39 10^{-44}~(sec.)
$$
The detailed analysis of this issue is out of the scope of the
present article.

There might be an interesting interconnection on how the discrete time
is used in information dynamics theory\cite{IKO,Ohya} and
the discrete time dynamics as it appears in presented study.
In particular, it would be interesting to consider the
equations for information dynamics with discrete time.

Finally, we would like to comment also that there might be a deep
interrelation between the energy-time uncertainty relations
\cite{MT,Rue,PfeFro} and Bohr-Somerfeld quantization rules
\cite{MasFed} in quantum mechanics and our discrete time model. In
particular one can try to write the Bohr-Somerfeld semi-classical
quantization rules for energy and time as canonical variables. For a
system with conserved energy one might get $E_n T_n \sim n\hbar$,
this relation holds for example for energy levels $E_n$ and
classical periods $T_n$ of Hydrogen atom. One the other hand the
relation (\ref{percd}) in discrete mechanics could be treated as the
condition for the period to take only discrete values $T_n\sim
n\tau$. We can see that although relations are similar there is an
extra factor $E_n$ in the quantum-mechanical relation. In fact one
may argue that if we make the $\tau$ in equations of motion depend
on the energy of the system as
$$
\tau=\tau_0 \frac{\varepsilon}{E},
$$
where $\tau_0$ is the ``fundamental'' time quantum and $\varepsilon$
a ``fundamental'' energy quantum, we get precisely the semiclassical
quantization rules. The question arise how to treat the energy $E$
here and what will happen with the dynamics. Further investigation
of this interrelation will be discussed elsewhere.

\section{Acknowledgments}

The authors would like to thank B.~Hiley, A.~Plotnitsky, G.~`t
Hooft, H.~Gus\-tafson, and K.~Gustafson for discussions on
quantum-like models with discrete time and M.~Ohya and D.~Petz for
discussions on classical and quantum information theory and information
dynamics.
Y.V. would like to thank L.~Joukovskaya for fruitful
discussions on atomic spectra measurements.

\end{document}